\documentclass[12pt,preprint]{aastex}
\begin{document}
\title{The Anisotropic Two-Point Correlation Functions of the Nonlinear 
Traceless Tidal Field in the Principal-Axis Frame}
\author{\sc Jounghun Lee\altaffilmark{1}, Oliver Hahn\altaffilmark{2}, and 
Cristiano Porciani\altaffilmark{2,3}}
\altaffiltext{1}{Department of Physics and Astronomy, FPRD, 
Seoul National University, Seoul 151-742 , Korea}
\altaffiltext{2}{Institute for Astronomy, ETH Zurich, CH-8093 Z\"urich, 
Switzerland}
\altaffiltext{3}{Argelander Institut f\"ur Astronomie, Auf dem H\"ugel 71, 
D-53121 Bonn, Germany}
\email{jounghun@astro.snu.ac.kr}
\begin{abstract}

Galaxies on the largest scales of the Universe are observed to be embedded in 
the filamentary cosmic web which is shaped by the nonlinear tidal field. 
As an efficient tool to quantitatively describe the statistics of 
this cosmic web, we present the anisotropic two-point correlation functions 
of the nonlinear traceless tidal field in the principal-axis frame, which are 
measured using numerical data from an N-body simulation. We show that 
both of the nonlinear density and traceless tidal fields are more strongly 
correlated along the directions perpendicular to the eigenvectors 
associated with the largest eigenvalues of the local tidal field. 
The correlation length scale of the traceless tidal field is found to be 
$\sim 20\, h^{-1}$Mpc, which is much larger than that of the density field 
$\sim 5\, h^{-1}$Mpc. We also provide analytic fitting formulae for the 
anisotropic correlation functions of the traceless tidal field, which turn 
out to be in excellent agreement with the numerical results. We expect that 
our numerical results and analytic formula are useful to disentangle 
cosmological information from the filamentary network of the large-scale 
structures.
\end{abstract}

\section{INTRODUCTION}
On the largest scales, the spatial distribution of galaxies exhibits the
pattern of a filamentary network which is commonly referred to as the 
"cosmic web". N-body simulations have long predicted the web-like pattern in 
the clustering of matter and galaxies \citep[e.g.,][]{sha-zel86}, 
recent large galaxy surveys have confirmed its presence \citep[e.g.,][]{2df}, 
and theoretical endeavors have been able to explain the 
naturalness of its occurrence in the context of the cold dark matter (CDM) 
paradigm \citep{bon-mye96,bon-etal96}. It is now well accepted that the 
cosmic web is a real, natural, and common phenomenon. 

Nonetheless, the most crucial part has been missing for the comprehension 
of the cosmic web - an optimal statistical tool to quantify it. 
The standard N-point statistics of course can be used in principle to 
quantify it, but they cannot capture the essence of the cosmic web that 
it is a nonlinear manifestation of the primordial tidal interactions. 
It is because in N-point statistics what is emphasized is the role of gravity 
in establishing the high-order correlations of the density field. 
The cosmic web is a nonlinear manifestation of the spatial coherence of
the tidal field \citep{bon-etal96}. Thus, to quantify the cosmic web 
and capture its essence, it is necessary to have a statistical tool by which 
the role and effect of the tidal field and its spatial correlation are 
highlighted. 

Patchy efforts have so far been made to account for the effect of the tidal 
field on structure formation. For instance, the failure of the Press-Schecther 
mass function \citep{ps74} has been attributed to the deviation of the 
gravitational process from spherical dynamics caused by the effect of the 
tidal forces \citep{mon95,aud-etal97,lee-sha98,she-etal01}. 
The galaxy angular momentum has been ascribed to the initial tidal 
interactions with the surrounding matter distribution 
\citep{dor70,whi84,cat-the96,lee-pen00,por-etal02a,por-etal02b}.
The intrinsic galaxy alignments are also found to be explained due to 
the effect of the nonlinear tidal interactions between the galaxies.
\citep{cat-etal01,cri-etal01,lee-etal05,lee-pen07,lee-pen08}. 
The alignment and ellipticity correlation functions have been developed 
to quantify the preferential orientations of galaxies relative to the 
surrounding large-scale structure \citep{fal-etal09}. 
The cosmic web, however, requires not a patchy but a unified way in which 
the multiple aspects of the nonlinear tidal field can be quantified.

The difficulty in quantifying the effect of the tidal field lies in the 
fact that the correlations of the tidal field are highly anisotropic even 
in the linear regime. \citet{lee-pen01} and \citet{cat-por01} have for the 
first time derived analytically the anisotropic two-point correlations of the 
linear tidal field. But, their results were all expressed in the fixed 
Cartesian coordinate frame. To fully appreciate the anisotropic nature of the 
correlations of the tidal field, it is desirable to find the correlations 
of the tidal field in the principal axis frame, since the filamentary 
cosmic web is generated by the anisotropic collapse of matter along the 
principal axes of the tidal field \citep{bon-etal96}. 
Unfortunately, it is extremely difficult to derive analytically the 
correlations of even the linear tidal field in the principal axis frame 
which rotates from point to point.

The goal of this Paper is to measure numerically the anisotropic two-point 
correlations of the nonlinear tidal field in the principal-axis frame and 
to provide an analytic fitting formula as an efficient statistical 
tool for the description of the cosmic web. This paper is organized as 
follows. In \S 2, we provide a brief description of the numerical data. 
In \S 3, we present the numerical results and analytic fitting formula 
as well. In \S 4, we summarize the results and discuss the implications.

\section{DATA}

We use the N-body results derived by \citet{hah-etal07} using the 
tree-PM code GADGET-2 \citep{spr-etal05a}. The simulations followed the 
evolution of $512^3$ particles starting from redshift $z=52.4$ to $z=0$ in 
a periodic box of linear size $180h^{-1}$Mpc. A flat $\Lambda$CDM 
cosmology is assumed with key parameters given as $\Omega_{m}=0.25$ 
(matter density), $\Omega_{\Lambda}=0.75$ (vacuum energy density ), 
$\Omega_{b}=0.045$ (baryon density), $\sigma_{8}=0.9$ 
(power spectrum amplitude), $H_{0}=0.73$ (Hubble constant) and 
$n_{s}=1$ (spectral index). 

\citet{hah-etal07} constructed the overdensity field $\delta$ (dimensionless 
density contrast) on $1024^{3}$ grids with the help of the Cloud-In-Cell 
interpolation method and smoothed it by a Gaussian kernel with a filtering 
radius $R_{\rm s}$. The peculiar gravitational potential field $\phi$ was 
also derived from the overdensity field by solving the Poisson's equation 
$\nabla^2\phi\propto \delta $. In simulations, the overdensity field is 
converted to the Fourier space with the help of a fast Fourier transform 
(FFT). The potential field is obtained as $\phi\propto k^{-2}\delta$ 
where $k$ represents the magnitude of the wave vector in the Fourier space.
For a detailed description of the N-body simulations and the derivation 
of the density and potential fields, see \citet{hah-etal07}. 
For the smoothing of the density (and potential) fields, we use the small 
smoothing scale $R_{s}=0.5h^{-1}$Mpc to minimize the effect of the smoothing 
on the determination of the correlations of the tidal fields.

\begin{figure}
  \begin{center}
   \plotone{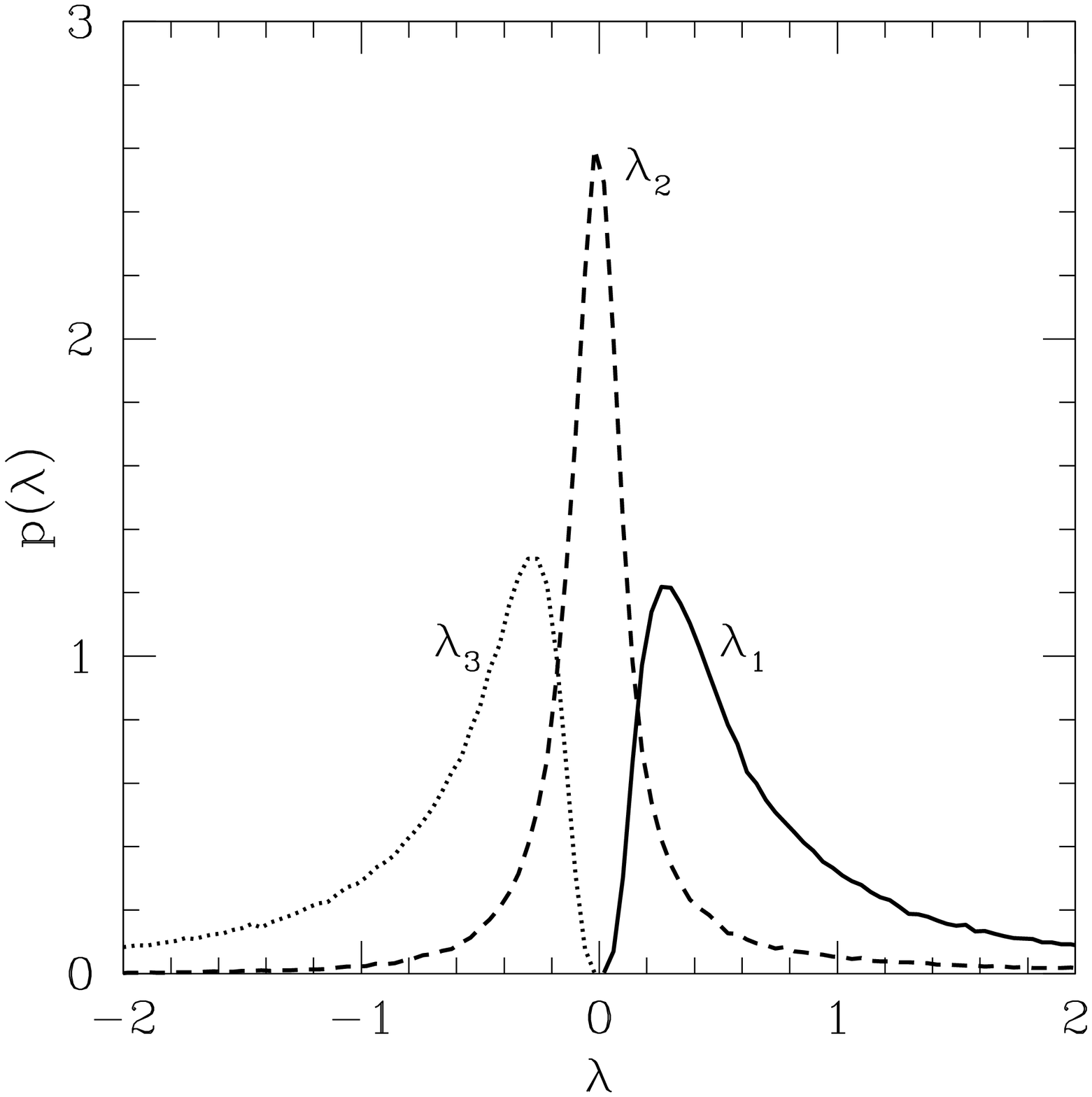}
\caption{Probability density distributions of the three eigenvalues, 
$\lambda_{1}$, $\lambda_{2}$ and $\lambda_{3}$ (solid, dotted and 
dashed line, respectively) of the traceless tidal field smoothed on 
the scale of $0.5\, h^{-1}$Mpc.}
\label{fig:pro}
 \end{center}
\end{figure}

The tidal tensor $(T_{ij})$ is defined as the Hessian of the peculiar 
gravitational potential $T_{ij}\equiv\partial_i\partial_j\,\phi\;$. 
In Fourier space, it is written as $T_{ij}= k_{i}k_{j}\,\phi\;$. 
Using the FFT again, we calculate the tidal tensor at each of the randomly 
selected $250000$ grid points. The sum of the diagonal components of 
$(T_{ij})$ (i.e., the trace of the tidal field) equals the dimensionless 
density contrast, $\delta$. To sort out the true tidal effect from the 
gravitational effect, we calculate the traceless tidal field 
$(\tilde{T}_{ij})$ as $\tilde{T}_{ij}\equiv T_{ij}-(\delta/3)I_{ij}$ where 
$(I_{ij})$ denotes the identity tensor. Diagonalizing $\tilde{T}_{ij}$ at 
each of the 250000 sampling points, we find its three eigenvalues, 
$\{\lambda_{1},\ \lambda_{2},\ \lambda_{3}\}$ (in a decreasing order, i.e., 
$\lambda_{1}>\lambda_{2}>\lambda_{3}$), and the corresponding normalized 
eigenvectors with unit magnitude, 
$\{\hat{\bf e}_{1},\ \hat{\bf e}_{2},\ \hat{\bf e}_{3}\}$.
Regardless of whether a given volume element is expanding or contracting 
(negative or positive $\delta$) the traceless tidal tensor $\tilde{T}_{ij}$ 
gives the deformation of that volume element (relative to spherical expansion 
or contraction). By construction, we have $\lambda_1 >0$ and $\lambda_3 <0$.  
The second largest eigenvalue $\lambda_2$ can take on either sign. 
The maximal (relative) compression of a given volume element by gravity 
occurs along the direction parallel to $\hat{\bf e}_{1}$ associated with 
$\lambda_1$. 
Figure \ref{fig:pro} plots the probability density distributions of 
$\lambda_{1},\ \lambda_{2}$ and $\lambda_{3}$ as solid, dashed and dotted 
lines, respectively. The probability density distribution of $\lambda_{1}$ 
and $\lambda_{3}$ is positively and negatively skewed, respectively while 
that of $\lambda_{2}$ is almost symmetric about zero. Note that the two 
distributions, $p(\lambda_{1})$ and $p(\lambda_{3})$, are also almost 
symmetric to each other around zero: $p(\lambda_{1})\sim p(-\lambda_{3})$.  
The distribution $p(\lambda_{2})$ is much narrower than $p(\lambda_{1}$).

In the following section, we measure the two-point correlation functions 
of $\lambda_{1},\ \lambda_{2},\ \lambda_{3}$ and $\delta$ in the 
principal-axis frame of the traceless tidal field and study their 
characteristic behaviors.
   
\section{NUMERICAL ANALYSIS AND FITTING FORMULA}

For each pair of points located at ${\bf x}$ and ${\bf x}+{\bf r}$, 
we express the separation vector ${\bf r}$ in the frame of the principal 
axes of $\tilde{T}_{ij}({\bf x})$ spanned by 
$\{\hat{\bf e}_{1},\ \hat{\bf e}_{2},\ \hat{\bf e}_{3}$\} with orienting 
the $z$-axis in the direction of $\hat{\bf e}_{1}$ (the first principal 
axis of the local tidal field). Let $(r,\theta,\phi)$ represent the spherical 
polar coordinate of ${\bf r}$ in the principal-axis frame. 
For every grid pair, we calculate the product, 
$\lambda_{i}({\bf x})\lambda_{i}({\bf x}+{\bf r})$ (for $i=1,\ 2,\ 3$). 
Then, we take the average of the product over the points whose separation 
vectors belong to the differential bin, $[{\bf r},{\bf r}+d{\bf r}]$, 
to determine the correlation of $\lambda_{i}$ as a function of 
$(r, \theta, \phi)$ as 
\begin{equation}
\label{eqn:anixi}
\xi_{\lambda i}(r,\theta,\phi) = 
\langle\lambda_{i}({\bf x})\cdot\lambda_{i}({\bf r}+{\bf x})\rangle.
\end{equation}

\begin{figure}
  \begin{center}
   \plotone{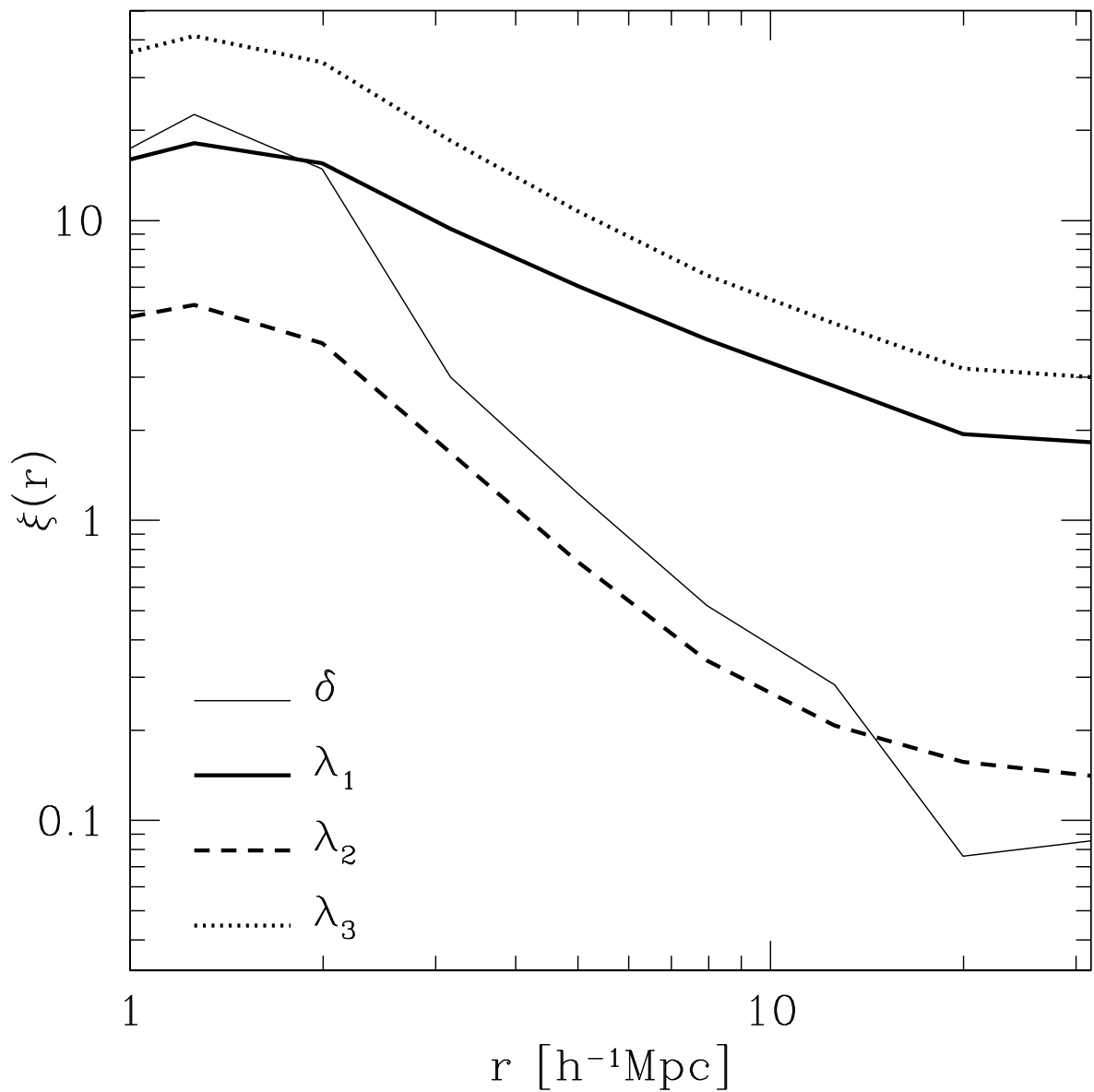}
\caption{Two-point isotropic correlation functions of the three eigenvalues 
of the traceless tidal field, $\lambda_{1}$, $\lambda_{2}$ and $\lambda_{3}$ 
(solid, dashed and dotted line, respectively) smoothed on the scale 
of $0.5h^{-1}$Mpc. The two-point isotropic correlations of $\delta$ is 
also plotted (thin solid line) for comparison.}
\label{fig:iso}
 \end{center}
\end{figure}

Figure \ref{fig:iso} shows the isotropic two-point correlations 
$\xi_{\lambda i}(r)$ of the three eigenvalues of the traceless tidal field 
as a function of $r$ (solid, dashed and dotted line for 
$\lambda_1,\ \lambda_2$ and $\lambda_3$, respectively) at $z=0$. 
This isotropic correlation function $\xi_{\lambda i}(r)$ is obtained by 
taking the average $\lambda_{i}({\bf x})\lambda_{i}({\bf x}+{\bf r})$ over 
the points whose separation distances belong to the differential distance 
bin, $[r,r+dr]$. 
The tendency of $\xi_{\lambda i}(r)$ being flat for $r<2\, h^{-1}$Mpc is due 
to the effect of the smoothing of the tidal field. The isotropic two-point 
correlation of $\delta$ is also plotted for comparison (thin solid line). 
As it can be seen, the correlations of $\lambda_{1}$ and $\lambda_{3}$ 
decrease with $r$ much more slowly than that of $\delta$. In other words, 
the largest and smallest eigenvalues of the traceless tidal field are coherent 
over larger scales than the density field $\delta$. Let the correlation length 
scale $l_{c}$ correspond to the distance at which the correlation strength 
becomes an order of magnitude lower than its value at $r=0$. We find 
that $l_{c}\approx 20h^{-1}$Mpc for $\lambda_{1}$ and $\lambda_{3}$, 
while $\delta$ and $\lambda_{2}$ has $l_{c}\approx 3h^{-1}$Mpc. 
The smallest eigenvalue $\lambda_{3}$ has the highest amplitude at all 
distances, while the second largest eigenvalue $\lambda_{2}$ has the lowest 
amplitude at $r\le 10h^{-1}$Mpc. The correlation amplitude of 
$\delta$ is comparable to that of $\lambda_{1}$ at small distance 
$r\le 2h^{-1}$Mpc but decreases much more rapidly with $r$ 
than $\lambda_{1}$, reaching the lowest value at $r\ge 10h^{-1}$Mpc.

\begin{figure}
  \begin{center}
   \plotone{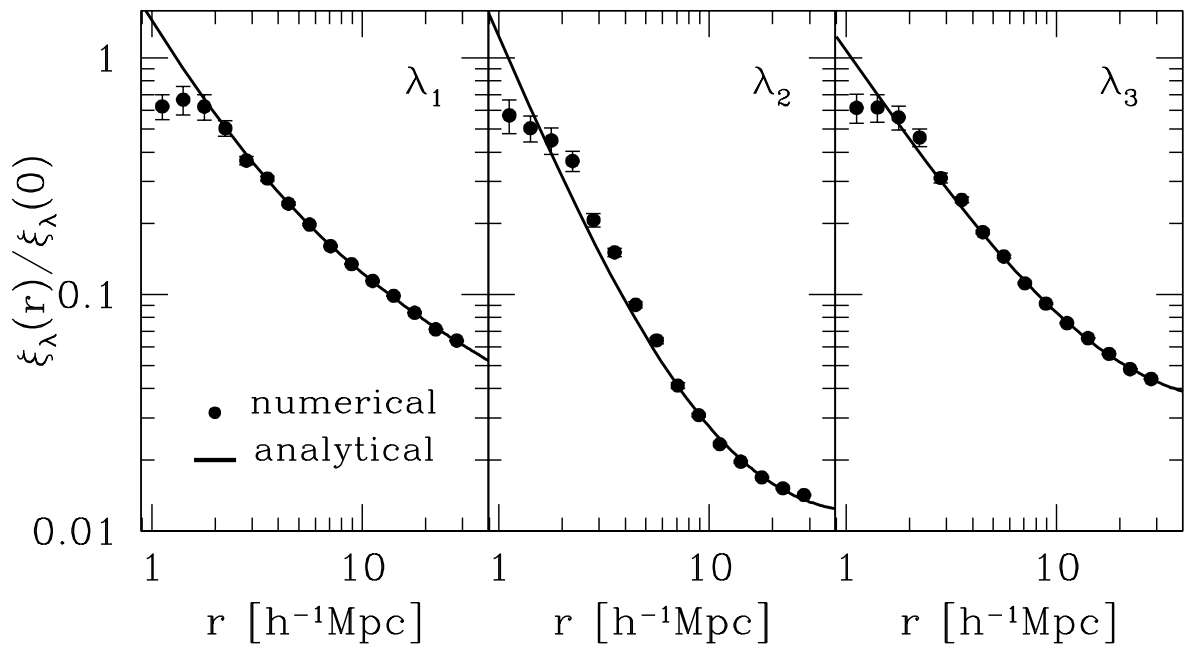}
\caption{Rescaled isotropic correlation functions of $\lambda_{1}$ (right), 
$\lambda_{2}$ (middle) and $\lambda_{3}$ (left) in the principal-axis 
frame of the local tidal field. In each panel the dots represent the numerical 
results while the solid line is the analytic fitting formula. The errors 
represent one standard deviation in the measurement of the mean values.}
\label{fig:lam}
 \end{center}
\end{figure}

For the rescaled isotropic correlation 
$\xi_{\lambda i}(r)/\xi_{\lambda i}(0)$, we find the following analytic 
formula:
\begin{eqnarray}
\label{eqn:isofit1}
\frac{\xi(r)}{\xi(0)} &=& 
\left(\frac{r}{r_{0}}+1\right)^{n}r^{n-3},\quad {\rm for} \quad 
\lambda_{1}, \lambda_{3}, \\
\label{eqn:isofit2}
\frac{\xi(r)}{\xi(0)} &=& 
\left(\frac{r}{r_{0}}+1\right)^{n}r^{1/2-n},\quad {\rm for} \quad 
\lambda_{2}
\end{eqnarray}
where $r_{0}$ and $n$ are two fitting parameters. By comparing the 
numerical results with the above analytic formula, we empirically determine
the best-fit values of $r_{0}$ and $n$, which are listed in Table 1.
Figure \ref{fig:lam} shows the rescaled isotropic correlations of 
$\lambda_{1}$ (left), $\lambda_{2}$ (middle) and $\lambda_{3}$ (right) at 
$z=0$. In each panel the dots and the solid line represent the numerical 
results and the analytic model with best-fit parameters, respectively. 
The error at each bin is calculated as one standard deviation in the 
measurement of the mean value. As it can be seen, the analytic models 
(eqs.[\ref{eqn:isofit1}-[\ref{eqn:isofit2}]) fit the numerical results very 
well for $r\ge 2\, h^{-1}$Mpc. On smaller distances $r< 2h^{-1}$Mpc, 
the numerical results become flatter than the analytic curve due to the 
smoothing effect.

\begin{figure}
  \begin{center}
   \plotone{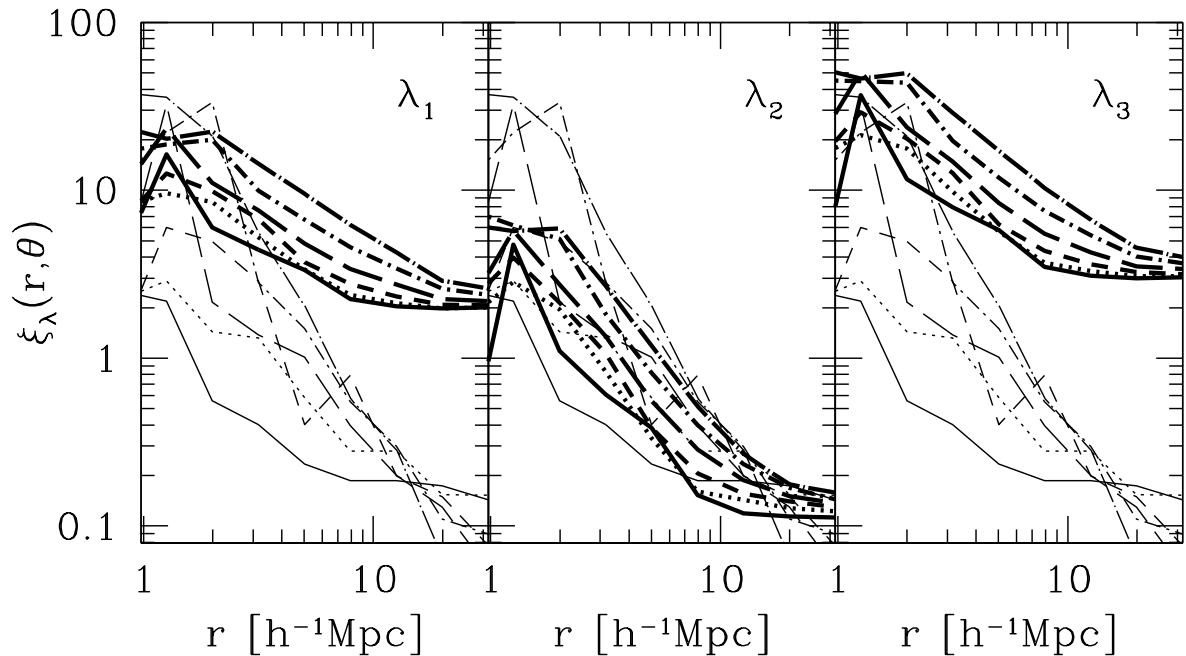}
\caption{Anisotropic correlation functions of $\lambda_{1}$ (right), 
$\lambda_{2}$ (middle) and $\lambda_{3}$ (left) as a function of $r$ and 
their variations with polar angle $\theta$ in the principal-axis frame of 
the local tidal field (thick solid lines). The polar angle $\theta$ 
represents the angle between the position vector and the first eigenvector 
of the local tidal field. In each panel the solid, dotted, dashed, 
long-dashed, dot-dashed, and dot-longdahsed line correspond to the range, 
$0^{\circ}\le \theta <15^{\circ}$, $15^{\circ}\le \theta <30^{\circ}$, 
$30^{\circ}\le \theta <45^{\circ}$, $45^{\circ}\le \theta <60^{\circ}$, 
$60^{\circ}\le \theta <75^{\circ}$ and $75^{\circ}\le \theta\le 90^{\circ}$, 
respectively. The anisotropic correlation functions  of $\delta$ are also 
shown for comparison (thin solid lines).}
\label{fig:ani}
 \end{center}
\end{figure}

Figure \ref{fig:ani} shows the anisotropic two-point correlations of 
$\lambda_{1}$ (left), $\lambda_{2}$ (middle) and $\lambda_{3}$ (right) 
as a function of $r$ and their variation with $\theta$ at $z=0$. 
These anisotropic correlations $\xi_{\lambda i}(r,\theta)$ are obtained 
by taking the average of $\lambda_{i}({\bf x})\lambda_{i}({\bf x}+{\bf r})$ 
over the points whose separation distances belong to the differential 
distance bin $[r,r+dr]$ and at which the angles between the separation vectors 
and the first principal axes are in the differential range of 
$[\theta,\theta+d\theta]$. 
In each panel the thick solid, dashed, dotted, longdashed, dot-dashed, 
and dot-longdashed line correspond to the range 
$0^{\circ}\le \theta <15^{\circ}$, $15^{\circ}\le \theta <30^{\circ}$, 
$30^{\circ}\le \theta <45^{\circ}$, $45^{\circ}\le \theta <60^{\circ}$, 
$60^{\circ}\le \theta <75^{\circ}$ and $75^{\circ}\le \theta\le 90^{\circ}$,  
respectively. The fluctuations of $\xi_{\lambda_i}(r,\theta)$ in the small-$r$ 
section ($r\le 2\, h^{-1}$Mpc) are due to the numerical noise. 
The anisotropic two-point correlation of $\delta$ is also plotted for 
comparison as thin lines in each panel. As it can be seen, both correlations 
of $\lambda$ and $\delta$ increase with $\theta$ in the principal-axis frame 
at all distances. That is, the correlations become stronger toward the 
directions normal to the first principal axes of the tidal field. 
This behavior is consistent with the scenario that matter compress maximally 
along the first principal axes of the tidal field and thus is more likely to 
lie in the plane normal to the first principal axis of the tidal field. 
The variation of $\lambda_{2}$ with $\theta$ is stronger than that of 
$\lambda_{1}$ and $\lambda_{3}$. Note also that the degree of variation of 
$\xi_{\lambda_{i}}(r,\theta)$ with $\theta$ decreases with $r$. 
In other words, the correlations tend to become less anisotropic at 
large distances ($r\ge 10\, h^{-1}$Mpc).

We also obtain $\xi_{\lambda_{i}}(\theta)$ by taking the average of 
$\lambda_{i}({\bf x})\lambda_{i}({\bf x}+{\bf r})$ over the points at which 
the angles between the first principal axis and the separation vectors belong 
to the differential angle bin, $[\theta,\theta+d\theta]$. The following 
analytic formulae are found to fit the numerical results well: 
\begin{eqnarray}
\label{eqn:anifit1}
\frac{\xi(\theta)}{\xi(0)} &=& \frac{1}{1-A\sin^{4}\theta},\quad 
{\rm for} \quad \lambda_{1}, \lambda_{3}, \\
\label{eqn:anifit2}
\frac{\xi(\theta)}{\xi(0)} &=& \frac{1}{1-A\sin^{2}\theta},\quad 
{\rm for} \quad 
\lambda_{2}
\end{eqnarray}
where $A$ is an adjustable parameter whose best-fit value is determined 
via the $\chi^{2}$ statistics (see Table 1). Figure \ref{fig:ani} shows 
the anisotropic correlation function of $\lambda_{1}$ (right), 
$\lambda_{2}$ (middle) and $\lambda_{3}$ (left) as a function of 
$\theta$ in the principal axis frame of the local tidal field. 
In each panel the dots represent the numerical results while the solid line 
is the analytic fitting formula. As it can be seen, the analytic models fit 
the numerical results very well for each case. To improve readibility we do 
not show errorbars which are negligibly small anyway. 
These results show clearly how anisotropic the correlations of the 
eigenvalues of the traceless tidal field are with respect to the first 
principal axis. The difference between the correlations at $\theta=0$ and 
$\theta=90^{\circ}$ reaches approximately $60\%$ for $\lambda_{1}$ and 
$\lambda_{3}$, while it increases up to $70\%$ for $\lambda_{2}$.

\section{SUMMARY AND DISCUSSION}

The web-like pattern in the large-scale matter distribution is an imprint 
produced by the effect of the nonlinear tidal field in the late universe. 
While the trace part of the tidal field is responsible for the collapse of 
matter, its traceless part deforms the cosmic volume elements, resisting 
the overall spherical expansion or contraction. In consequence the traceless 
tidal field induces anisotropy in the large-scale correlations of matter 
distribution. An efficient statistical tool has long been looked for to 
describe quantitatively this cosmic web and to retrieve cosmological 
information encoded in the cosmic web.

We have determined numerically the two-point correlations of the three 
eigenvalues of the nonlinear traceless tidal field defined as the Hessian 
of the gravitational potential in the frame of the principal axes of the tidal 
field. The numerical findings indicate that the correlation functions of the 
traceless tidal field and the density field are all anisotropic relative to 
the principal axes. Their correlations having much larger correlation length 
scales than that of the density field. increase along the directions normal 
to the first principal axes of the tidal field.  
The analytic fitting formula for the correlation 
functions of the three eigenvalues of the tidal field are determined and 
found to fit the numerical results well. Our numerical and analytical 
results are thus able to provide a physical description of the large-scale 
filamentary structure. 

It will be interesting to know how the anisotropic two-point correlations 
of thetraceless tidal field depend on the background cosmology.
Our future work is in this direction.

\acknowledgments

J.L. acknowledges the financial support from the Korea Science and 
Engineering Foundation (KOSEF) grant funded by the Korean Government 
(MOST, NO. R01-2007-000-10246-0). O.H. acknowledges support from the 
Swiss National Science Foundation. All simulations were performed on the
Gonzales cluster at ETH Zurich, Switzerland.

\clearpage

\clearpage
\clearpage
\clearpage
\clearpage
 \begin{figure}
  \begin{center}
   \plotone{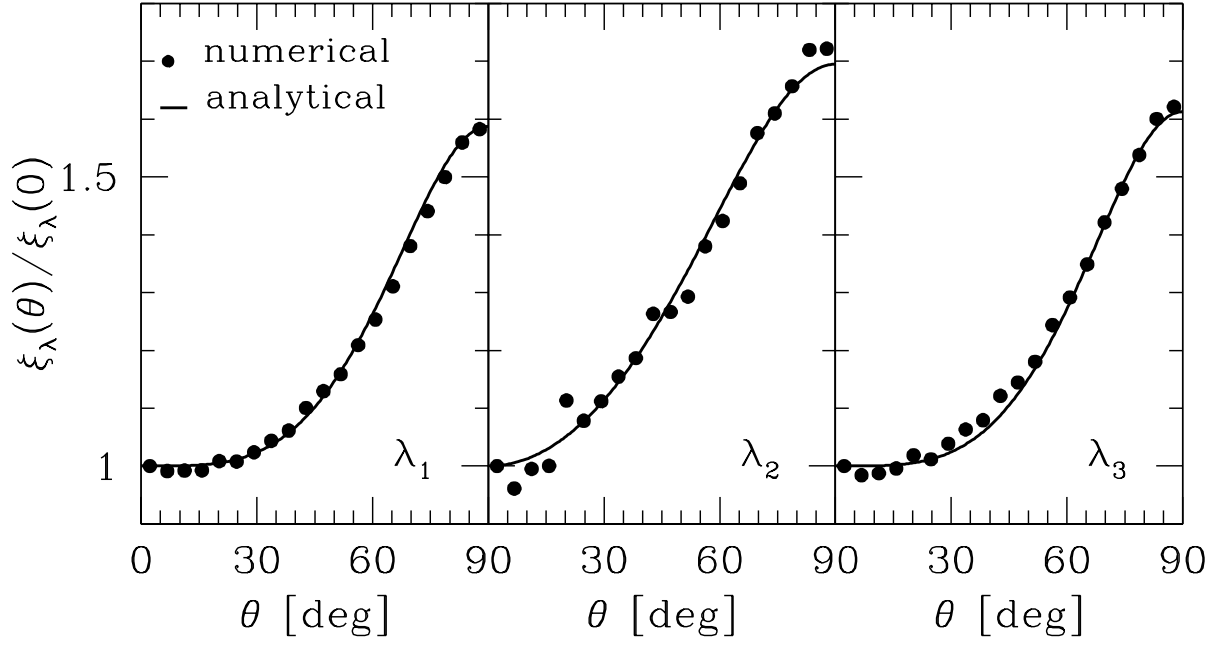}
\caption{Anisotropic correlation function of $\lambda_{1}$ (right), 
$\lambda_{2}$ (middle) and $\lambda_{3}$ (left) as a function of 
$\theta$ in the principal axis frame of the local tidal field. 
In each panel the dots represent the numerical results while the solid line 
is the analytic fitting formula.}
\label{fig:lam1}
 \end{center}
\end{figure}
\clearpage
\begin{deluxetable}{cccc}
\tablewidth{0pt}
\setlength{\tabcolsep}{5mm}
\tablehead{parameters & $\lambda_{1}$  & $\lambda_{2}$ & $\lambda_{3}$}
\tablecaption{Best-fit values of the parameters in the fitting formula 
for the two-point anisotropic correlation functions of the three eigenvalues 
of the nonlinear traceless tidal field.}
\startdata   
$n$ &  $1.28$ & $2.76$ & $1.65$  \\
$r_{0}$ & $3.03$ & $12.58$ & $21.51$  \\
$A$ & $0.37$ & $0.41$ & $0.38$  \\
\enddata
\label{tab:bestfit}
\end{deluxetable}
\end{document}